\documentclass[twocolumn,english,aps,prl,showpacs]{revtex4}
\usepackage[T1]{fontenc}
\usepackage[latin1]{inputenc}
\usepackage{amsmath}
\usepackage{graphicx}
\usepackage{amssymb}
\usepackage{esint}

\makeatletter
\@ifundefined{textcolor}{}
{%
 \definecolor{BLACK}{gray}{0}
 \definecolor{WHITE}{gray}{1}
 \definecolor{RED}{rgb}{1,0,0}
 \definecolor{GREEN}{rgb}{0,1,0}
 \definecolor{BLUE}{rgb}{0,0,1}
 \definecolor{CYAN}{cmyk}{1,0,0,0}
 \definecolor{MAGENTA}{cmyk}{0,1,0,0}
 \definecolor{YELLOW}{cmyk}{0,0,1,0}
 }

\@ifundefined{definecolor}{\@ifundefined{definecolor}
 {\usepackage{color}}{}
}{}\makeatother

\makeatother

\usepackage{babel}

\makeatother

\usepackage{babel}

\makeatother

\usepackage{babel}

\makeatother

\usepackage{babel}

\begin{document}

\title{Manipulating Majorana fermions in one-dimensional spin-orbit coupled
atomic Fermi gases}

\author{Xia-Ji Liu$^{1}$ and P. D. Drummond$^{1}$ }

\affiliation{$^{1}$ARC Centre of Excellence for Quantum-Atom Optics, Centre for
Atom Optics and Ultrafast Spectroscopy, Swinburne University of Technology,
Melbourne 3122, Australia}

\date{\today}
\begin{abstract}
Majorana fermions are promising candidates for storing and processing
information in topological quantum computation. The ability to control
such individual information carriers in trapped ultracold atomic Fermi
gases is a novel theme in quantum information science. However, fermionic
atoms are neutral and thus are difficult to manipulate. Here, we theoretically
investigate the control of emergent Majorana fermions in one-dimensional
spin-orbit coupled atomic Fermi gases. We discuss (i) how to move
Majorana fermions by increasing or decreasing an effective Zeeman
field, which acts like a solid state control voltage gate; and (ii)
how to create a pair of Majorana fermions by adding a magnetic impurity
potential. We discuss the experimental realization of our control
scheme in an ultracold Fermi gas of $^{40}$K atoms. 
\end{abstract}


\maketitle
Majorana fermions - particles that are their own antiparticles - were
proposed by Ettore Majorana in 1937 \cite{Majorana1937}. They are
finding ever wider relevance in modern physics \cite{Wilczek2009}.
A well-known proposal is the use of Majorana fermions as information
carriers in topological quantum information processing \cite{Kitaev2006,Nayak2008}.
Due to the robustness of topological quantum computation, research
into the creation and manipulation of Majorana fermions has become
an active research topic in a variety of fields of physics, ranging
from condensed matter physics to ultracold atomic physics. Most theoretical
schemes of processing topological quantum information based on Majorana
fermions have relied on using solid-state systems \cite{Fu2008,Alicea2011}.
Evidence of the existence of Majorana fermions in hybrid superconductor-semiconductor
nanowire devices has been reported very recently \cite{Mourik2012}.

The realization of Majorana fermions in ultracold atomic systems has
also been discussed extensively \cite{Tewari2007,Zhang2008,Sato2009,Jiang2011,Liu2012a,Liu2012b}.
Compared with solid state systems, a unique advantage of ultracold
atoms is their unprecedented controllability and purity, which may
be expected to lead to reduced decoherence. A vast range of interactions,
geometries and dimensions are possible: using the tool of Feshbach
resonances, one can control the interatomic interactions very accurately.
By using the technique of optical lattices, one can create artificial
one-dimensional (1D) or two-dimensional (2D) environments to explore
how physics changes with dimensionality. In this context, topological
quantum computation based on Majorana zero-energy modes in the vortex
core of a 2D atomic Fermi gas has been discussed \cite{Tewari2007}.
However, it is very difficult to experimentally generate and manipulate
a vortex lattice in a fermionic superfluid.

In this Brief Report, we theoretically investigate the manipulation
of Majorana fermions in one-dimensional spin-orbit coupled atomic
Fermi gases. The existence of Majorana fermions in such systems was
recently checked by us, by performing fully microscopic calculations
within the mean-field Bogoliubov-de Gennes (BdG) theory \cite{Liu2012b}.
Here, we study how to control Majorana fermions using a background
magnetic field combined with a magnetic impurity potential. Our investigation
is motived by the recent realization of a three-dimensional spin-orbit
coupled Fermi gas of $^{40}$K and $^{6}$Li atoms, at ShanXi University
\cite{exptShanXi} and at MIT \cite{exptMIT}, respectively. We anticipate
that a 1D spin-orbit coupled Fermi gas will be created very soon,
by using 2D optical lattices \cite{Liao2010}. At the end of this
paper, we discuss the possible experimental realization of our control
scheme for $^{40}$K atoms.

\textbf{Model Hamiltonian}. - Our theoretical approach is outlined
in detail in the previous work \cite{Liu2012b}. Here we extend this
approach to include a classical magnetic impurity. In the following,
we introduce briefly the essential ingredients of the theory. The
model Hamiltonian of a 1D atomic Fermi gas with spin-orbit coupling,
as realized at ShanXi \cite{exptShanXi} and at MIT \cite{exptMIT},
can be written as \begin{eqnarray}
{\cal H} & = & \int dx\psi^{\dagger}\left(x\right)\left[{\cal H}_{\sigma}^{S}\left(x\right)-h\sigma_{z}+\lambda k\sigma_{y}\right]\psi\left(x\right)\nonumber \\
 &  & +g_{1D}\int dx\psi_{\uparrow}^{\dagger}\left(x\right)\psi_{\downarrow}^{\dagger}\left(x\right)\psi_{\downarrow}\left(x\right)\psi_{\uparrow}\left(x\right),\label{Hami}\end{eqnarray}
 where $\psi^{\dagger}\left(x\right)\equiv[\psi_{\uparrow}^{\dagger}\left(x\right),\psi_{\downarrow}^{\dagger}\left(x\right)]$
denote collectively the creation field operators for spin-up and spin-down
atoms, $h$ denotes the strength of an effective magnetic field causing
a Zeeman splitting, while $\lambda k\sigma_{y}\equiv-i\lambda(\partial/\partial x)\sigma_{y}$
is the spin-orbit coupling term with coupling strength $\lambda$.
In addition, $\sigma_{y}$ and $\sigma_{z}$ are the usual $2\times2$
Pauli matrices while $g_{1D}=-2\hbar^{2}/(ma_{1D})$ is the effective
Hamiltonian coupling given an $s$-wave scattering length in one dimension
of $a_{1D}$. We note that the existence of spin-orbit coupling, magnetic
fields and quantum transport to the cloud edge remove all of the usual
symmetry properties of Fermi gases, and these systems therefore provide
an opportunity to investigate the least symmetric of the nonstandard
symmetry classes in normal-superconducting hybrid structures\cite{Altland}.

The single-particle Hamiltonian ${\cal H}_{\sigma}^{S}(x)$ describes
the atomic motion in a harmonic trapping potential $m\omega^{2}x^{2}/2$,
together with a scattering potential of a classical magnetic impurity,
modeled as $-V_{imp}\left(x\right)\sigma_{z}$. Combining these terms,
${\cal H}_{\sigma}^{S}(x)$ takes the form, \begin{equation}
{\cal H}_{\sigma}^{S}(x)=-(\hbar^{2}/2m)(\partial^{2}/\partial x^{2})+(m/2)\omega^{2}x^{2}-V_{imp}(x)\sigma_{z}-\mu,\label{eq:sphami-1}\end{equation}
 where $\mu$ is the chemical potential. The impurity scattering potential
with strength $V_{0}$ is taken to have a gaussian shape, so that
\begin{equation}
V_{imp}(x)=[V_{0}/(\sqrt{2\pi}d)]\exp[-(x-x_{0})^{2}/(2d^{2})],\end{equation}
 where $x_{0}$ and $d$ are the position and width of the impurity
potential, respectively. Thus, in the limit of infinitely narrow width
where $d\rightarrow0$, we obtain a Dirac delta-function like potential,
$V_{imp}\left(x\right)=V_{0}\delta(x-x_{0})$.

We solve the model Hamiltonian (\ref{Hami}) using a mean-field Bogoliubov-de
Gennes (BdG) approach. The wave-function of low-energy quasiparticles
$[u_{\uparrow\eta}\left(x\right),u_{\downarrow\eta}\left(x\right),v_{\uparrow\eta}\left(x\right),v_{\downarrow\eta}\left(x\right)]^{T}$
with energy $E_{\eta}$ satisfies the BdG equation, \begin{widetext}
\begin{equation}
\left[\begin{array}{cccc}
{\cal H}_{\uparrow}^{S}(x)-h & -\lambda\partial/\partial x & 0 & -\Delta(x)\\
\lambda\partial/\partial x & {\cal H}_{\downarrow}^{S}(x)+h & \Delta(x) & 0\\
0 & \Delta^{*}(x) & -{\cal H}_{\uparrow}^{S}(x)+h & \lambda\partial/\partial x\\
-\Delta^{*}(x) & 0 & -\lambda\partial/\partial x & -{\cal H}_{\downarrow}^{S}(x)-h\end{array}\right]\left[\begin{array}{c}
u_{\uparrow\eta}\left(x\right)\\
u_{\downarrow\eta}\left(x\right)\\
v_{\uparrow\eta}\left(x\right)\\
v_{\downarrow\eta}\left(x\right)\end{array}\right]=E_{\eta}\left[\begin{array}{c}
u_{\uparrow\eta}\left(x\right)\\
u_{\downarrow\eta}\left(x\right)\\
v_{\uparrow\eta}\left(x\right)\\
v_{\downarrow\eta}\left(x\right)\end{array}\right],\label{BdG}\end{equation}
 \end{widetext} where the pairing field is defined as:\begin{align}
\Delta(x)= & -(g_{1D}/2)\sum_{\eta}[u_{\uparrow\eta}(x)v_{\downarrow\eta}^{*}(x)f(E_{\eta})+\nonumber \\
 & +u_{\downarrow\eta}(x)v_{\uparrow\eta}^{*}(x)f(-E_{\eta})]\end{align}
 and the density of spin-$\sigma$ atoms is given by $n_{\sigma}\left(x\right)=(1/2)\sum_{\eta}[\left|u_{\sigma\eta}(x)\right|^{2}f(E_{\eta})+\left|v_{\sigma\eta}(x)\right|^{2}f(-E_{\eta})]$.
Here, $f\left(E\right)\equiv1/[e^{E/(k_{B}T)}+1]$ is the Fermi distribution
function at temperature $T$. The BdG equation (\ref{BdG}) can be
solved self-consistently using a hybrid method detailed in Ref. \cite{Liu2007}. 

In the following calculations we consider a typical dimensionless
interaction parameter of $\gamma\equiv\left(a_{ho}/a_{1D}\right)/\left(\pi N^{1/2}\right)=1.6$,
where $a_{ho}\equiv\sqrt{\hbar/(m\omega)}$ is the characteristic
oscillator length of the trap and $N=100$ is the total number of
fermionic atoms. For spin-orbit coupling, we take $\lambda k_{F}/E_{F}=1$,
where $k_{F}$ and $E_{F}$ are respectively the Fermi wave-vector
and Fermi energy. The unit of length is the Thomas-Fermi radius, $x_{F}=N^{1/2}a_{ho}$.
All calculations will be carried out at zero temperature, as the inclusion
of a finite but small temperature will only marginally affect our
results.

\textbf{A brief summary of previous results}. - In the {\em absence}
of magnetic impurity, the spin-orbit coupled 1D Fermi gas in a harmonic
trap will locally enter an interesting topological phase when the
Zeeman field satisfies $h>h_{c}(x)$, as we reported earlier in Ref.
\cite{Liu2012b}. Here $h_{c}(x)=\sqrt{\mu^{2}(x)+\Delta^{2}(x)}$
with $\mu\left(x\right)=\mu-m\omega^{2}x^{2}/2$ is the local critical
field at position $x$. For interaction strength $\gamma=1.6$ and
spin-orbit coupling $\lambda k_{F}/E_{F}=1$ mentioned above, this
occurs at $h>h_{c}\sim0.35E_{F}$ See, for example, the phase diagram
in Fig. 1. At higher magnetic fields of $h>h_{c}$, a mixed phase
emerges, consisting of a standard BCS superfluid at the trap center
and a topological superfluid at the two edges of the trap (see Fig.
2).

\begin{figure}[b]
\begin{centering}
\includegraphics[clip,width=0.4\textwidth]{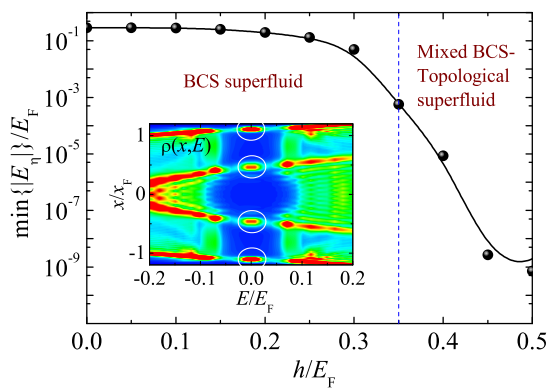} 
\par\end{centering}

\caption{(color online) Phase diagram determined from the lowest energy of
Bogoliubov quasiparticle spectrum. The inset shows the contour plot
of local density of state at $h/E_{F}=0.5$. The contributions from
Majorana fermions are highlighted by circles.}

\label{fig1} 
\end{figure}

As a salient feature of topological superfluids, Majorana fermions
appear at the edges. These Majorana fermions are nothing but the zero-energy
solutions of the BdG equation. Because of the particle-hole symmetry
of the BdG equation, the field operator of Bogoliubov quasiparticles
satisfies $\Gamma_{E}=\Gamma_{-E}^{\dagger}$. If the energy $E=0$,
we would then have $\Gamma_{0}=\Gamma_{0}^{\dagger}$, which is the
defining feature of Majorana fermions. Related to this zero-energy
feature, it is easy to see that spatially resolved radio-frequency
(r.f.) spectroscopy, which measures the local density of state of
the atoms, defined as $\rho(x,E)=(1/2)\sum_{\sigma\eta}[\left|u_{\sigma\eta}(x)\right|^{2}\delta(E-E_{\eta})+\left|v_{\sigma\eta}(x)\right|^{2}\delta(E+E_{\eta})]$,
provides a useful means to identify Majorana fermions. In the inset
of Fig. 1, we can identify from r.f. spectroscopy four zero-energy
Majorana fermions at the Zeeman field $h=0.5E_{F}$, two for each
topological superfluid.

\textbf{Manipulating Majorana fermions}. - Let us consider how to
manipulate Majorana fermions in 1D atomic Fermi gas. We focus on two
operations: to move and to create Majorana fermions.

\begin{figure}[htp]
\begin{centering}
\includegraphics[clip,width=0.4\textwidth]{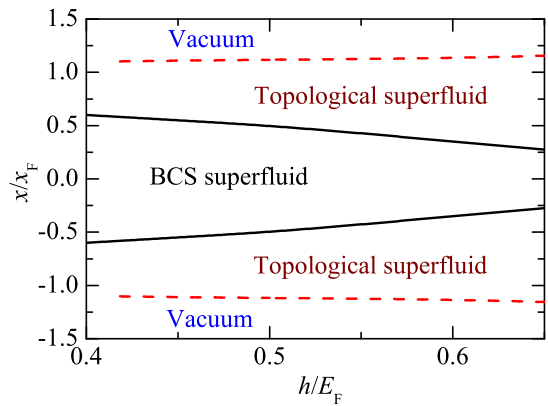} 
\par\end{centering}

\caption{(color online) Positions of Majorana fermions as a function of Zeeman
field.}

\label{fig2} 
\end{figure}

We propose to use the Zeeman field, as an analog to a voltage gate
in solid-state systems \cite{Alicea2011}, to transport Majorana fermions
through the whole Fermi cloud. Indeed, the position of Majorana fermions
depends critically on the Zeeman field, as shown in Fig. 2, where
we plot the boundary of each superfluid phase. The four Majorana fermions
sit exactly on these boundaries. Therefore, by {\em adiabatically}
increasing or decreasing the Zeeman field, the inner two Majorana
fermions will move towards or away from the trap center. The position
of the outer two Majorana fermions, however, is less affected by the
Zeeman field.

To create new Majorana fermions in the system, we may consider either
(i) creating a new topological superfluid in some areas of the Fermi
cloud or (ii) introducing a strong defect in the existing topological
superfluids. In the former case, a new pair of Majorana fermions will
appear at the edges of the induced new topological superfluid. In
the latter case, the strong defect will create a low-density hole,
which is basically vacuum and thus topologically trivial. New Majorana
fermions will then be trapped at the hole center. A well-known example
of the latter case is a vortex in 2D topological superfluid, which
hosts a zero-energy Majorana fermion mode inside the vortex core.

The main result of this work is that these two ideas (i) and (ii)
of creating new Majorana fermions can be realized by introducing a
classical magnetic impurity. The magnetic impurity serves as an attractive
or repulsive scattering potential for spin-up or spin-down atoms,
respectively, as we mentioned earlier in the model Hamiltonian. Due
to the finite width of the impurity scattering potential, the first
idea (i) seems more practical. In the following, we will focus on
this idea and examine it by using self-consistent BdG calculations.
To test this concept, we will put an impurity at the center of the
Fermi cloud (i.e., $x_{0}=0$) and take a potential width $d=0.07x_{F}$.
The scattering strength of the impurity is taken as $V_{0}=0.13x_{F}E_{F}$.

\begin{figure}[b]
\begin{centering}
\includegraphics[clip,width=0.45\textwidth]{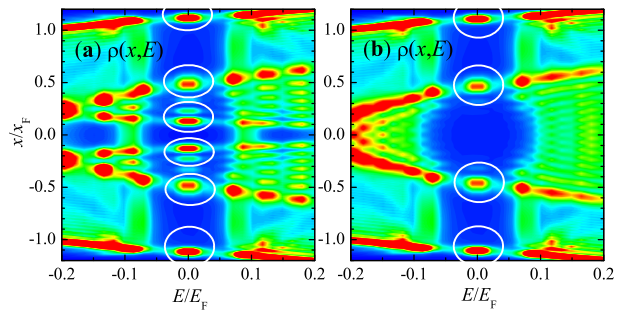} 
\par\end{centering}

\caption{(color online) Linear contour plot of the local density of states
with (a) or without (b) a magnetic impurity at $h/E_{F}=0.5$.}

\label{fig3} 
\end{figure}

\begin{figure}[htp]
\begin{centering}
\includegraphics[clip,width=0.4\textwidth]{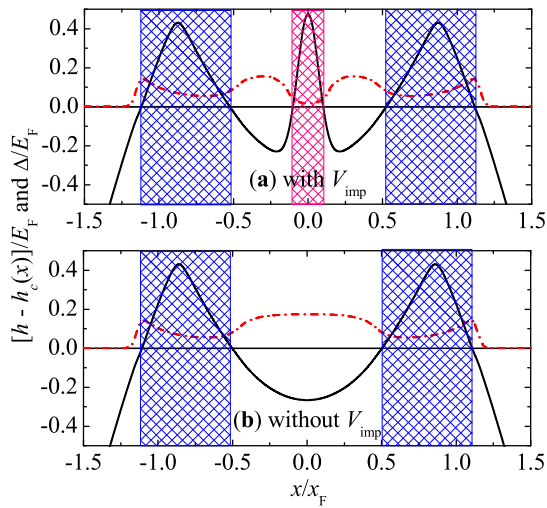} 
\par\end{centering}

\caption{(color online) Criterion for topological superfluid $h-h_{c}(x)$
(solid lines) and the superfluid order parameter $\Delta(x)$ (dot-dashed
lines) at $h/E_{F}=0.5$. The upper and lower panels show the results
with and without a magnetic impurity, respectively. The cross-patterns
highlight the areas of topological superfluids.}

\label{fig4} 
\end{figure}

The existence of impurity induced Majorana fermions is clearly evident
from the local density of state, which is measurable using spatially
resolved r.f. spectroscopy, as reported in Fig. 3a using a linear
contour plot. For comparison, we show also in Fig. 3b the expected
spectroscopy scan without an impurity. It is clear that a new pair
of Majorana fermions is created at about $x=\pm0.1x_{F}$, while other
four Majorana fermions almost keep the same positions as those without
the impurity potential. The creation of new Majorana fermions can
be easily understood. The magnetic impurity potential with a relatively
broader potential width may be regarded as a local Zeeman field. By
tuning the strength of the magnetic impurity, we can greatly enhance
the local Zeeman field at a particular area. This creates a local
topological superfluid that hosts new Majorana fermions at its edges.
To check this picture, we have calculated the criterion for a local
topological superfluid, $h-h_{c}(x)>0$. As highlighted in Fig. 4a
by a pink cross-pattern, we find that this criterion is indeed satisfied
at about the origin, where the magnetic impurity is located. The superfluid
gap is significantly reduced at the magnetic impurity.

\begin{figure}[htp]
\begin{centering}
\includegraphics[clip,width=0.4\textwidth]{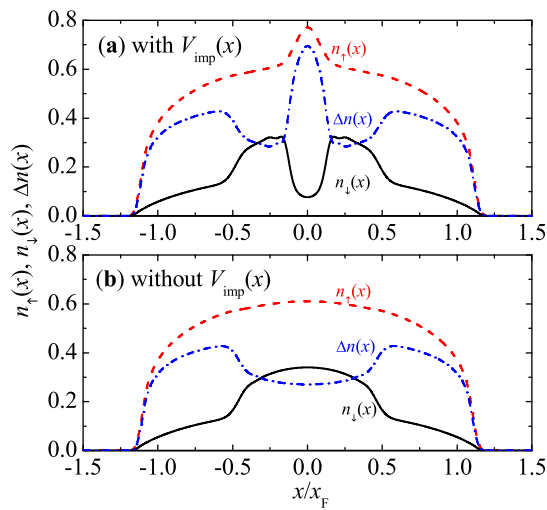} 
\par\end{centering}

\caption{(color online) The spin-up and spin-down density distribution, $n_{\uparrow}$
(dashed lines) and $n_{\downarrow}$ (solid lines), and their difference
$\Delta n=n_{\uparrow}-n_{\downarrow}$ (dot-dashed lines) at $h/E_{F}=0.5$.}

\label{fig5} 
\end{figure}

We finally check how the density distribution is affected by magnetic
impurity. In Fig. 5a, we present the spin-up and spin-down density
distributions and their difference $\Delta n(x)=n_{\uparrow}(x)-n_{\downarrow}(x)$.
Compared with the result without the impurity (Fig. 5b), we find that
the local density is enhanced and reduced for the spin-up and spin-down
atoms, respectively, due to the magnetic impurity. This is consistent
with the picture of inducing a local Zeeman field, as mentioned in
the above.

\textbf{Experimental proposal}. - The model Hamiltonian Eq. (\ref{Hami})
can be realized using ultracold $^{40}$K atoms \cite{exptShanXi}.
The two spin-1/2 states can be chosen as the two magnetic sublevels
$\left|\uparrow\right\rangle =\left|F=9/2,m_{F}=-9/2\right\rangle $
and $\left|\downarrow\right\rangle =\left|F=9/2,m_{F}=-7/2\right\rangle $
of the $F=9/2$ hyperfine level. To create the spin-orbit coupling,
one can use a pair of counter-propagating Raman laser beams with strength
$\Omega_{R}$, which flip atoms from $\left|\downarrow\right\rangle $
to $\left|\uparrow\right\rangle $ spin states and simultaneously
impart momentum $2\hbar k_{R}$ (or vice versa), via the two-photon
Raman process \cite{exptShanXi}. This is the same technique used
in the National Institute of Standards and Technology (NIST) to generate
synthetic spin-orbit coupling in a Bose-Einstein condensate of $^{87}$Rb
atoms \cite{Lin2011}. 

In our model Hamiltonian, the spin-orbit coupling strength is related
to the wave-vector $k_{R}$ by $\lambda=\hbar^{2}k_{R}/(2m)$, and
the effective Zeeman field can be tuned by the strength of Raman laser
beam, i.e., $h=\Omega_{R}/2$. To make the system one-dimensional,
one can impose 2D optical lattices that restrict atoms moving along
a single tube. This was already demonstrated at Rice University to
obtain an imbalanced 1D Fermi gas \cite{Liao2010}. It is possible
to create an impurity potential by using a laser beam that has a sufficient
narrow beam width, as schematically shown in Fig. 6. By suitably tuning
its frequency, the scattering potential caused by the laser beam can
be attractive for the spin-up state and repulsive for the spin-down
state. This localized probe therefore behaves like a classical magnetic
impurity. All the techniques required to simulate the model Hamiltonian
Eq. (\ref{Hami}) are therefore within the reach of current experiments. 

We note that our scheme does not yet lead to a fully developed quantum
gate, which in any event requires a careful analysis of quantum dynamics
during the switching process. Instead, it provides a simple demonstration
system whereby the concepts of creating and moving Majorana fermions
in a quasi one-dimensional environment - like `birds on a wire' -
can be usefully investigated.

\begin{figure}[b]
\begin{centering}
\includegraphics[clip,width=0.3\textwidth]{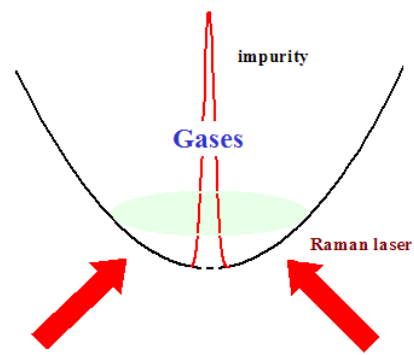} 
\par\end{centering}

\caption{(color online) Schematic plot of the experimental setup. The spin-orbit
coupling is induced by a pair of Raman laser beams, while the magnetic
impurity with finite width is created by a dimple laser beam.}

\label{fig6} 
\end{figure}

\textbf{Conclusions}. - In summary, we have theoretically proposed
how to manipulate Majorana fermions in a one-dimensional spin-orbit
coupled atomic Fermi gas in a harmonic trap. The transportation of
Majorana fermions through the Fermi cloud may be controlled by an
effective Zeeman field, while the creation of pairs of Majorana fermions
can be achieved through an artificial magnetic impurity created using
a dimple laser beam. Our proposal may be viewed as the cold-atom analog
of the control gates proposed in the solid-state systems \cite{Alicea2011}.
We anticipate that these control operations can be realized using
ultracold $^{40}$K atoms \cite{exptShanXi}.

We thank Hui Hu and Han Pu for useful discussions. This work is supported
by the ARC Discovery Project (Grant No. DP0984637) and the NFRP-China
(Grant No. 2011CB921502).


\begin{thebibliography}{19}
\bibitem{Majorana1937} E. Majorana, Nuovo Cimento \textbf{14}, 171
(1937).

\bibitem{Wilczek2009} F. Wilczek, Nat. Phys. \textbf{5}, 614 (2009).

\bibitem{Kitaev2006} A. Kitaev, Ann. Phys. (NY) \textbf{321}, 2 (2006).

\bibitem{Nayak2008} C. Nayak, S. Simon, A. Stern, M. Freedman, and
S. Das Sarma, Rev. Mod. Phys. \textbf{80}, 1083 (2008).

\bibitem{Fu2008} L. Fu and C. L. Kane, Phys. Rev. Lett. \textbf{100},
096407 (2008).

\bibitem{Alicea2011} J. Alicea, Y. Oreg, G. Refael, F. von Oppen
and M. P. A. Fisher, Nat. Phys. \textbf{7}, 412 (2011).

\bibitem{Mourik2012} V. Mourik, K. Zuo, S. M. Frolov, S. R. Plissard,
E. P. A. M. Bakkers, and L. P. Kouwenhoven, Science \textbf{336},
1003 (2012).

\bibitem{Tewari2007} S. Tewari, S. Das Sarma, C. Nayak, C. Zhang
and P. Zoller, \textbf{98}, 010506 (2007).

\bibitem{Liu2012b} X.-J. Liu and H. Hu, Phys. Rev. A \textbf{85},
033622 (2012).

\bibitem{Zhang2008} C. Zhang, S. Tewari, R. Lutchyn, and S. Das Sarma,
Phys. Rev. Lett. \textbf{101}, 160401 (2008).

\bibitem{Sato2009} M. Sato, Y. Takahashi, and S. Fujimoto, Phys.
Rev. Lett. \textbf{103}, 020401 (2009).

\bibitem{Jiang2011} L. Jiang, T. Kitagawa, J. Alicea, A. R. Akhmerov,
D. Pekker, G. Refael, J. I. Cirac, E. Demler, M. D. Lukin, and P.
Zoller, Phys. Rev. Lett. \textbf{106}, 220402 (2011).

\bibitem{Liu2012a} X.-J. Liu, L. Jiang, H. Pu, and H. Hu, Phys. Rev.
A \textbf{85}, 021603(R) (2012).

\bibitem{exptShanXi} P. Wang, Z.-Q. Yu, Z. Fu, J. Miao, L. Huang,
S. Chai, H. Zhai and J. Zhang, Phys. Rev. Lett. \textbf{109}, 095301
(2012).

\bibitem{exptMIT} L. W. Cheuk, A. T. Sommer, Z. Hadzibabic, T. Yefsah,
W. S. Bakr, and M. W. Zwierlein, Phys. Rev. Lett. \textbf{109}, 095302
(2012).

\bibitem{Liao2010} Y. A. Liao, A. S. C. Rittner, T. Paprotta, W.
Li, G. B. Partridge, R. G. Hulet, S. K. Baur, and E. J. Mueller, Nature
(London) \textbf{467}, 567 (2010).

\bibitem{Altland}Alexander Altland and Martin R. Zirnbauer, Phys.
Rev. B \textbf{55,} 1142 (1997).

\bibitem{Liu2007} X.-J. Liu, H. Hu, and P. D. Drummond, Phys. Rev.
A \textbf{76}, 043605 (2007).

\bibitem{Lin2011} Y.-J. Lin, K. Jiménez-Garc\'{i}a, and I. B. Spielman,
Nature (London) \textbf{471}, 83 (2011); X.-J. Liu et al, Phys. Rev.
Lett.\textbf{ 102}, 046402 (2009)
\end{thebibliography}
\end{document}